\documentclass[%
 reprint,
 superscriptaddress,
 amsmath,amssymb,
 aps,
floatfix,
]{revtex4-2}
\usepackage{float}
\usepackage[caption=false]{subfig}     
\usepackage[utf8]{inputenc}
\usepackage{natbib}
\usepackage{graphicx}
\usepackage{xcolor}
\usepackage{booktabs}
\usepackage{multirow}
\usepackage[utf8]{inputenc}
\newcommand{\ket}[1]{\left | #1 \right\rangle}

\usepackage{braket}
\usepackage{qcircuit}
\usepackage{hyperref}
\usepackage{tikz}
\usepackage{capt-of}

\begin{document}

\title{Multi-round QAOA and advanced mixers on a trapped-ion quantum computer}

\author{Yingyue Zhu}
\affiliation{Joint Quantum Institute and Department of Physics, University of Maryland, College Park, MD 20740, USA }
\author{Zewen Zhang}
\affiliation{Department of Physics and Astronomy, Rice University, Houston, TX 77005, USA }
\author{Bhuvanesh Sundar}
\affiliation{JILA, Department of Physics, University of Colorado, Boulder, CO 80309, USA}
\affiliation{Center for Theory of Quantum Matter, University of Colorado, Boulder, CO 80309, USA}
\author{Alaina M. Green}
\affiliation{Joint Quantum Institute and Department of Physics, University of Maryland, College Park, MD 20740, USA }
\author{C. Huerta Alderete}
\affiliation{Joint Quantum Institute and Department of Physics, University of Maryland, College Park, MD 20740, USA }
\author{Nhung H. Nguyen}
\affiliation{Joint Quantum Institute and Department of Physics, University of Maryland, College Park, MD 20740, USA }
\author{Kaden R. A. Hazzard} 
\affiliation{Department of Physics and Astronomy, Rice University, Houston, TX 77005, USA }
\affiliation{Rice Center for Quantum Materials, Rice University, Houston, TX 77005, USA }
\author{Norbert M. Linke}
\affiliation{Joint Quantum Institute and Department of Physics, University of Maryland, College Park, MD 20740, USA }
\affiliation{Duke Quantum Center and Department of Physics, Duke University, Durham, NC 27708, USA}

\date{\today}

\begin{abstract}
Combinatorial optimization problems on graphs have broad applications in science and engineering. The Quantum Approximate Optimization Algorithm (QAOA) is a method to solve these problems on a quantum computer by applying multiple rounds of variational circuits. However, there exist several challenges limiting the real-world applications of QAOA. In this paper, we demonstrate on a trapped-ion quantum computer that QAOA results improve with the number of rounds for multiple problems on several arbitrary graphs. We also demonstrate an advanced mixing Hamiltonian that allows sampling of all optimal solutions with predetermined weights. Our results are a step towards applying quantum algorithms to real-world problems.
\end{abstract}

\maketitle

\section{\label{sec:1}Introduction}
    Combinatorial optimization problems on graphs are ubiquitous in fields of science and engineering, such as bioinformatics~\cite{bio1,bio2}, earth science~\cite{earth_science}, logistics~\cite{green_logistics}, resource management~\cite{resource_management}, telecommunications~\cite{telecom}, e-commerce~\cite{e-commerce1, e-commerce2} and others. Efficient classical algorithms for solving many of these problems are not known, and quantum computers can potentially provide an advantage. The Quantum Approximate Optimization Algorithm (QAOA) has been used in several demonstrations to solve combinatorial as well as other types of optimization problems~\cite{PhysRevApplied.14.034010,google,photonics_implementation,benchmarkingQAOA_SC,XYinteraction,lao20212qan, PRXQuantum.1.020304,pagano_quantum_2020,otterbach2017unsupervisedML}. QAOA is a quantum-classical hybrid algorithm that produces high-quality approximate solutions~\cite{farhi2014quantum}. Even though it does not always guarantee an advantage over classical algorithms, QAOA can achieve a provable quadratic speedup in oracle calls when it is equivalent to Grover's algorithm~\cite{speedup1}, and numerical evidence shows that it can provide a polynomial speedup in some problems~\cite{QAOA_speedup2}. It has also been argued that even the output distribution from a one-round QAOA circuit is hard to sample classically~\cite{farhi2019quantum}. QAOA is also able to generate approximate answers with low-depth circuits, which makes it valuable for implementations on near-term quantum devices.

    In QAOA, two non-commuting Hamiltonians, the problem-dependent Hamiltonian $H_A$ and the mixing Hamiltonian $H_B$, are applied repeatedly to a chosen initial state $\ket{\psi_{\text{initial}}}$ in a bang-bang protocol. The final output state is an approximate ground state of $H_A$ as well as a solution to the optimization problem. 
     \begin{equation}\label{equ1}
        \ket{\psi_{\text{final}}}=\prod_{i=1}^{p}{e^{-iH_{B}\beta_{i}}e^{-iH_{A}\alpha_{i}}}\ket{\psi_{\text{initial}}}
    \end{equation}
    $\alpha_{i}$ and $\beta_{i}$ are real variational parameters, and $p$ is the number of QAOA rounds. The variational parameters are optimized classically to minimize the expectation value $\langle\psi_\textrm{final}|H_{A}|\psi_\textrm{final}\rangle$. $\ket{\psi_{\text{initial}}}$ is the ground state of $H_{B}$. In standard QAOA, $H_{B}$ is the $n$-qubit transverse-field Hamiltonian
     \begin{equation}\label{HB}
         H_{B}^{\text{transverse}}=\sum_{i=1}^{n}\sigma^{x}_{i},
    \end{equation} 
    where $\sigma^x_i$ is the Pauli X matrix acting on qubit $i$. 
    
    In theory, QAOA performance improves as $p$ is increased. However, increasing $p$ can degrade the results in practice when the experimental errors introduced by deeper circuits outcompete the theoretical QAOA gain. Additionally, if the connectivity of the graphs does not match the qubit connectivity of the quantum hardware, the overheads required to map these nonnative graphs to the qubits also greatly increase the circuit depth. Moreover, the standard QAOA often provides only a subset of the ground states, while many applications require knowledge about all of them. 
    
   The first result of our work is to show that the probability of finding a ground state with standard QAOA increases with $p$, up to $p=3$, on a trapped-ion quantum computer for optimization problems defined on arbitrary graphs.  Previous experimental works have demonstrated QAOA results improving with $p$ on hardware-native graphs~\cite{PhysRevApplied.14.034010,PRXQuantum.1.020304,lao20212qan,google}, while real-world graph problems are often hardware-nonnative.

    The second result of our work is to demonstrate that employing advanced mixing Hamiltonians in QAOA can allow one to access a broader range of classically hard optimization problems. The recently proposed Grover mixer QAOA (G-QAOA) \cite{sundar2019quantum,9259965} is capable of generating a superposition of all ground states with probabilities determined by their weights, which are defined in the optimization problem and provided as inputs. This feature is referred to as fair sampling. G-QAOA can be applied to both unweighted and weighted graph problems. The $n$-qubit Grover mixer takes the form
    \begin{equation}\label{Grover}
        H^{\text{Grover}}_{B}=\prod^{n}_{i=1}{\frac{1+(1-2q)\sigma^{z}_{i}+2\sqrt{q(1-q)}\sigma^{x}_{i}}{2}},
    \end{equation}
    where $q$ is related to the numerical weight assigned to each qubit, with $q=0.5$ corresponding to unweighted problems. Other important tasks relying on fair sampling include satisfiability (SAT)-based membership filters~\cite{FSapplication-SATmembfilter,FSapp-SAT,FSapp-SAT_annealing}, proportional model sampling~\cite{model_sampling}, machine learning~\cite{machinelearning1,machinelearning2}, and sampling the ground states of arbitrary classical spin Hamiltonians~\cite{classicalcounting2006,ROTH1996273}. Although in theory G-QAOA fairly samples ground states at any $p$, the total probability of finding ground states increases with $p$. While previous works have experimentally demonstrated one round of G-QAOA in Hamiltonian optimization problems on unweighted graphs~\cite{golden2021qaoabased,pelofske2021sampling}, we apply G-QAOA to both weighted and unweighted graph problems up to $p=2$ on arbitrary graphs, and quantitatively evaluate the experimental fair sampling results.    

    The experiments are implemented on a programmable trapped-ion quantum computer, where up to five $^{171}$Yb$^+$ ions in a linear chain are used as qubits. The qubit states $|0\rangle$ and $|1\rangle$ are encoded in the two hyperfine ground states $\ket{F=0,m_F=0}$ and $\ket{F=1,m_F=0}$ in the $^{2}\text{S}_{1/2}$ manifold. The qubits are initialized in $|0\rangle$ by optical pumping and read out with state-dependent fluorescence. Quantum controls are implemented by coherently manipulating the qubit states with two counter-propagating Raman beams, one of which is split into individual beams to address each qubit separately (see Appendix~\ref{A} for more details). 

    \begin{table*}[t] 
    \centering
    \begin{tabular}{c c l l p{1.6cm}|c c l l p{1cm}}
        \hline
        \multicolumn{5}{c|}{standard QAOA}&\multicolumn{5}{c}{G-QAOA}\\
        \hline
        \hspace{0.3cm}graph\hspace{0.5cm}&\multicolumn{1}{c}{}& $p=1$\hspace{0.7cm} &$p=2$\hspace{0.7cm} & $p=3$&\hspace{0.5cm}graph\hspace{0.5cm}&\multicolumn{1}{c}{}& $p=1$& \hspace{0.3cm}$p=2$& \\ 
        \hline
        \multirow{2}{*}{\hspace{0.3cm}triangle}  & sim\hspace{0.5cm}& 0.968 &1&& \multirow{2}{*}{\hspace{0.5cm}triangle}  & sim&0.781 &\hspace{0.3cm}0.999&\\
        &exp\hspace{0.5cm}&  0.943(2)&0.986(1)&& &exp & 0.739(2) &\hspace{0.3cm}0.751(2)&\\
        \cline{1-10}
        \multirow{2}{*}{\hspace{0.3cm}square}  & sim\hspace{0.5cm}&0.966&1&&\multirow{2}{*}{\hspace{0.5cm}square}  & \hspace{0.5cm}sim\hspace{0.5cm} &0.770 &\hspace{0.3cm}0.801&\\
        &exp &  0.931(1)& 0.949(1)& &&\hspace{0.5cm}exp\hspace{0.5cm} &  0.668(1)&\hspace{0.3cm}0.692(1)&\\ 
        \cline{1-10}
        \multirow{2}{*}{\hspace{0.3cm}paw}  & \hspace{0.5cm}sim\hspace{0.5cm}  & 0.871&0.958&0.985&\multirow{2}{*}{\hspace{0.5cm}paw}  & \hspace{0.5cm}sim\hspace{0.5cm}  & 0.645 &\hspace{0.3cm}0.867&\\
        &exp  &0.812(2)&0.819(1)&0.743(2)& & exp\hspace{0.5cm}& 0.548(2) &\hspace{0.3cm}0.551(2)&\\
        \hline
        \multirow{2}{*}{\hspace{0.3cm}bridge}  &\hspace{0.5cm}sim\hspace{0.5cm}  & 0.330 &0.557&0.996&\multirow{2}{*}{\hspace{0.5cm}paw}  & \hspace{0.5cm}sim\hspace{0.5cm}  & 0.105 &\hspace{0.3cm}0.835&\\
        &\hspace{0.5cm}exp\hspace{0.5cm}  & 0.300(3) &0.424(4)&0.660(4)&\hspace{0.5cm}(weighted)&\hspace{0.5cm}exp\hspace{0.5cm}  &0.235(1) &\hspace{0.3cm}0.421(1)&\\ 
        \hline
        \multicolumn{5}{c|}{}  &\multirow{2}{*}{\hspace{0.5cm}square}  &\hspace{0.5cm}sim\hspace{0.5cm} & 0.310 &\hspace{0.3cm}0.758&\\
        \multicolumn{5}{c|}{}  &\hspace{0.6cm}(weighted)&\hspace{0.5cm}exp\hspace{0.5cm} &  0.255(1)& \hspace{0.3cm}0.497(1)&\\ 
        \hline
        \hline
    \end{tabular}    
    \caption{\label{table1_v2}Simulated and experimental probabilities of finding a ground state for different $p$ by standard QAOA (left half) and G-QAOA (right half) for the edge cover (triangle, paw, and square graph) and Max-Cut problems (bridge graph). The errors are the statistical $68.3\%$ confidence interval.}
    \end{table*}
    
    \begin{figure}[ht]
        \centering
        \includegraphics[width=\linewidth]{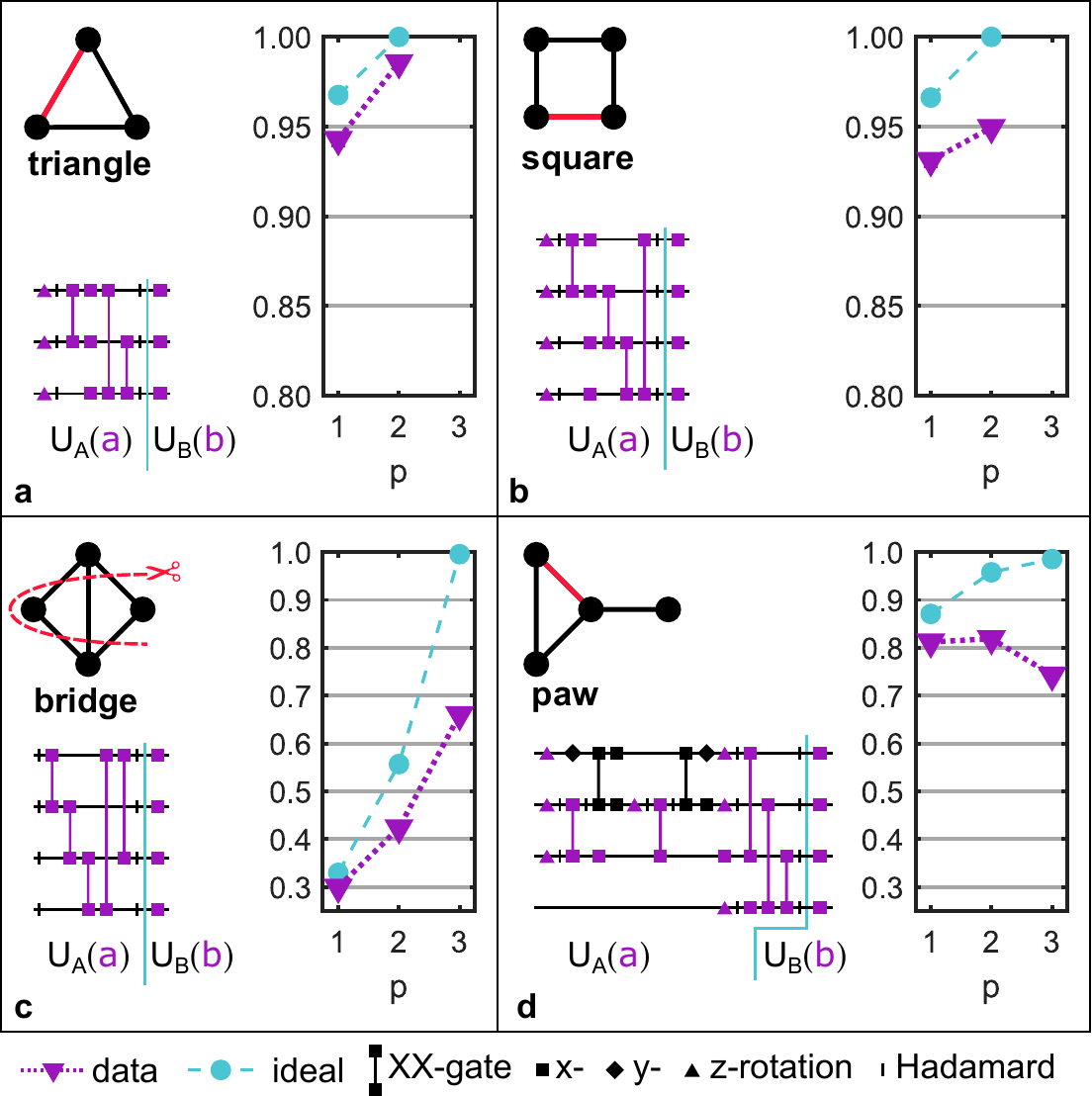} 
        \caption{Simulated and experimental results from standard QAOA for the edge cover and Max-Cut problems. The plots give the probability to find a solution on the trapped-ion machine compared to the ideal result for different numbers of QAOA rounds $p$. The graphs are indicated in the top left corner, i.e. the triangle (a), square (b), bridge (c) and paw graph (d). For edge cover problems (a,b,d) the black links show  example solutions, for the Max-Cut problem (c) the red line indicates the unique cut. Circuits show one round of the problem unitary $U_A=e^{-iH_{A}\alpha}$ and the mixer unitary $U_B=e^{-iH_{B}\beta}$, with purple gates parameterized by $\alpha$ or $\beta$. The values of parameters $\alpha_{i}$ and $\beta_{i}$ are listed in Appendix~\ref{B}. All qubits are initialized in the $x$-basis and measured in the $z$-basis (not shown). Statistical error bars on the data are smaller than the symbols. The dashed lines are a guide to the eye. }
        \label{fig:standardmixer}
    \end{figure}

\section{\label{sec:2}Higher-round QAOA on graph problems}
    In this section, we focus on finding a \emph{maximum cut} on the bridge graph and an \emph{edge cover} on the triangle, paw and square graphs with standard QAOA.  All graphs discussed in this section are unweighted. A graph $G$ is defined by a set of vertices $v\in V$, $|V|=N_{v} $,  and a set of edges $e\in E$, $|E|=N_{e} $.  A \emph{maximum cut} is a partition of all vertices into two complementary sets where the number of edges between them is maximized. An \emph{edge cover} is a subgraph $G'\subseteq G$ in which every $v \in V$ is connected to at least one edge $E'$ included in $G'$.
    A graph can have multiple \emph{maximum cuts} or \emph{edge covers}.

   When mapping the graphs onto a quantum computer, a qubit can represent either an edge or a vertex. In Max-Cut problems, the qubits encode the vertices. Each computational basis state represents a partition of $V$ with all vertices corresponding to qubits in state $|0\rangle$ in one set and the rest in the other set. The probability of finding each ground state is the population of the corresponding quantum state in $\ket{\psi_{\text{final}}}$. The problem Hamiltonian $H_{A}$ for Max-Cut problems is
    \begin{equation}\label{equ3}
        H_{A}^{\text{maxcut}}=\sum_{(i,j)\in E}^{}\sigma^{z}_{i}\sigma^{z}_{j}, 
    \end{equation}
    where $\sigma^z$ is the Pauli Z operator. If $(i,j)$ is an edge between sets, $\sigma^{z}_{i}\sigma^{z}_{j}=-1$. 

    In an edge cover problem, each qubit encodes a unique edge $e\in E$. Therefore, each computational basis state encodes a unique $G'$ where the qubit state $|0\rangle$ ($|1\rangle$) means that the corresponding edge is included (not included) in $G'$. $H_{A}$ encoding the edge cover problem is
    \begin{equation}\label{equ2}
    H_{A}^{\text{ec}}=\sum_{v\in V}\prod_{e\in E'(v)}\frac{1-\sigma^{z}_{e}}{2},
    \end{equation}
    where $E'(v)$ is the set of edges incident on vertex $v$ in the subgraph $G'$.
    
    In the experiment, the system is first prepared in the ground state of $H_B$ in Eq.~\eqref{HB}, $\ket{\psi_{\text{initial}}}=\ket{++...+}$ with $\ket{+}= \frac{1}{\sqrt{2}}(\ket{0}+\ket{1})$, by applying a Hadamard gate to each qubit. Then the system unitarily evolves under $H_{A}$ and ${H_{B}}$ alternately before being measured in the computational basis.
    
    The results are shown in Fig.~\ref{fig:standardmixer} and in Table~\ref{table1_v2} (left column). In the Max-Cut problem on the bridge graph, the probability of finding a \emph{maximum cut} clearly improves with $p$ [Fig.~\ref{fig:standardmixer}(c)], despite the $p$-fold increase in the number of gate operations. Similarly, the probabilities of finding one \emph{edge cover} on the triangle and square graph also increase with $p$ [Fig.~\ref{fig:standardmixer}(a,b)].
    
    However, on the paw graph, the probability of finding an \emph{edge cover} only improves marginally for $p=2$ and drops for $p=3$  [Fig.~\ref{fig:standardmixer}(d)]. In this problem, the implementation of $H^{\text{ec}}_{A}$ requires seven two-qubit entangling gates, the most among all four problems, and the additional gate error outweighs the theoretical gain with increasing $p$. The question then arises whether there are alternatives to increasing $p$ that will improve the solution probability. One such idea is to use more sophisticated mixers, an example of which is the Grover-mixer given in Eq.~\eqref{Grover}. While, as we will see, this does not increase the solution probability for a fixed gate depth, it does enable the solution of a new class of graph optimization problem—sampling problems.

    \begin{table*}[t]
        \centering
        \begin{tabular}{p{1.4cm} p{1.3cm} p{1.9cm} p{1.6cm} p{1.6cm} p{1.6cm} p{1.6cm} p{1.6cm} p{1.6cm}}
        \hline
            &&\hspace{2mm}$(N_{g})$       &2          &3          &4          &5         &6           &7\\
            \hline
            \multirow{4}{*}{triangle} &\multirow{2}{*}{$p=1$} &QAOA    &2.58   &5.00   &10.16   &           &           &\\
            &               &G-QAOA                                    &3.15   &5.87   &11.28   &           &           &\\
            &\multirow{2}{*}{$p=2$}  &QAOA                             &10.43   &26.48  &60.50(1)    &           &          &\\
            &   &G-QAOA                                                &3.11   &5.82   &11.30(2)   &           &           &\\
            \hline
            \multirow{4}{*}{paw}&\multirow{2}{*}{$p=1$} &QAOA          &3.13   &6.15    &11.65   &33.34   &            &\\
            &   &G-QAOA                                                &4.12   &7.19    &11.91   &21.54   &           &\\
            &\multirow{2}{*}{$p=2$}&QAOA                               &6.44    &14.93   &29.20(6)   &66.0(1)    &           &\\
            &   &G-QAOA                                                &4.09  &7.20(1)    &11.93   &21.61(3)   &           &\\
            \hline
            \multirow{4}{*}{square}&\multirow{2}{*}{$p=1$} &QAOA       &2.53   &4.28  &6.55  &9.78    &15.05   &25.09\\
            &       &G-QAOA                                            &3.16   &5.22   &7.84   &11.4(1)	   &16.82   &28.12\\
            &\multirow{2}{*}{$p=2$}&QAOA                               &6.64	   &15.33	&28.41 &53.7(1)	   &171.1(5)   &493(1)\\
            &   &G-QAOA                                                &3.25   &5.36	&7.99	&11.56  &16.87   &27.69\\
            \hline
            \hline
        \end{tabular}
        \caption{\label{counting}Number of draws required to see $N_{g}$ ground states on unweighted graphs. Average and error bars are calculated from 100,000 independent trials. All error bars less than 0.01 are not included in the table. In comparison, if this similar counting test is done on random guessing ($p=0$), it will take 16.64(3), 36.58(6) and 41.40(4) of draws in these three graphs to get all ground states.}
    \end{table*}

\section{\label{sec:3}Sampling with G-QAOA on unweighted graphs}
     
    All four unweighted graph problems studied in Sec.~\ref{sec:2} have more than one solution, but the standard QAOA favors only one. G-QAOA samples all ground states with equal probability at any $p$ on unweighted graphs. In this section, the goal is to find all solutions for the three edge cover problems studied in Sec.~\ref{sec:2} by using G-QAOA. $H_{A}^{\text{ec}}$ is given in Eq.~\eqref{equ2}. But now $H_{B}$ is the Grover mixer for unweighted graphs given in Eq.~\eqref{Grover} with $q=0.5$. Circuits for implementing the Grover mixer can be found in Fig.~\ref{fig:circuit3qGrover} and Fig.~\ref{fig:circuit4qGrover} in Appendix~\ref{B}.
  
    \begin{figure}
        \centering
        \includegraphics[width=\linewidth]{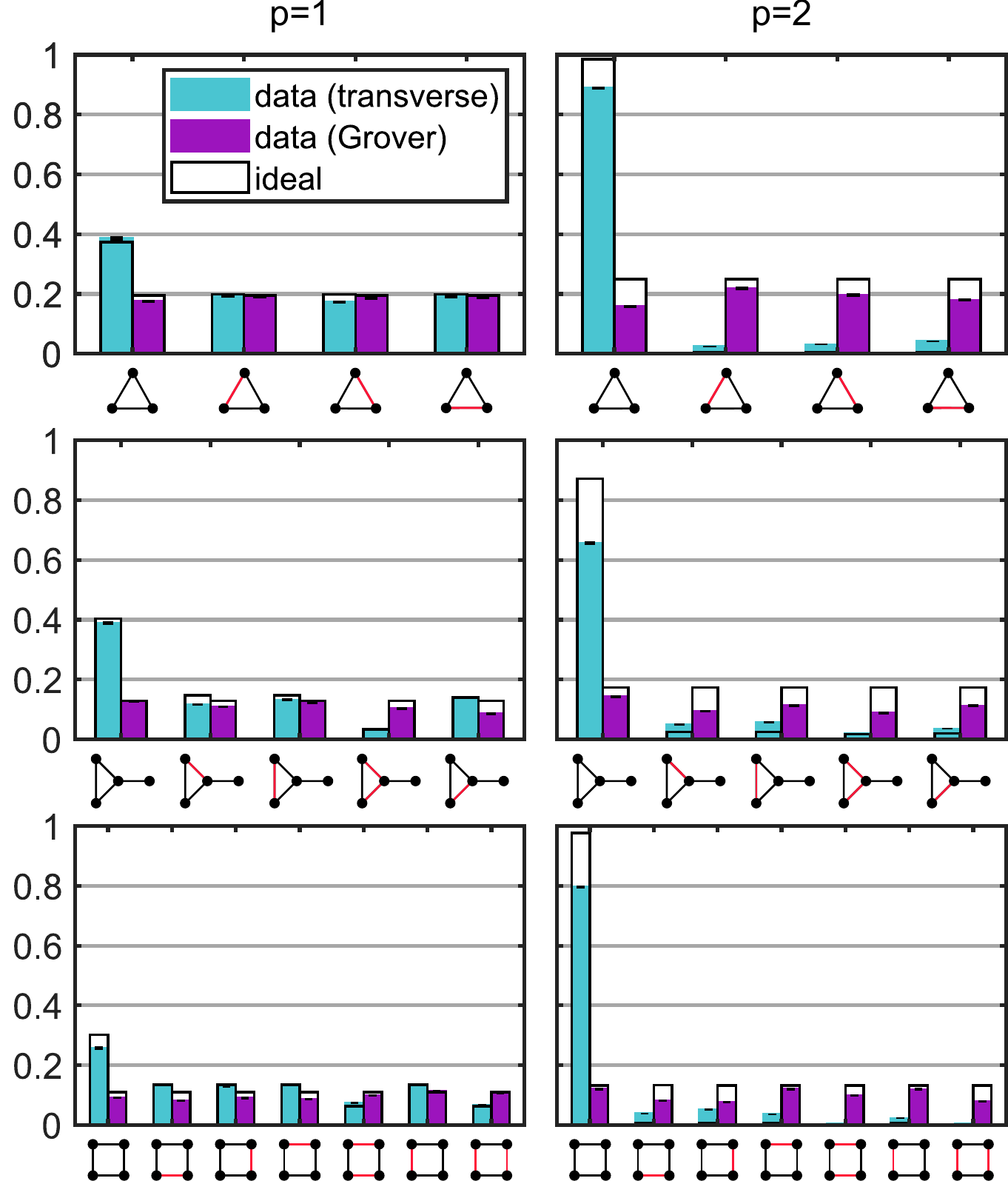}
        \caption{Simulated and experimental probabilities of finding each ground state by standard QAOA and G-QAOA for the edge cover problems on the triangle, paw, and square graph. Each plot compares the individual probabilities of finding each ground state at the same $p$ for the same problem between standard QAOA and G-QAOA. Different \emph{edge covers} are illustrated at the bottom of each plot on the horizontal axis, with edges included in each specific \emph{edge cover} colored in black and edges not included colored in red. Simulated probabilities are delineated in contour with solid black lines, overlapping with corresponding experimental data plotted with colored bar without any outline. Error bars are shown as short black dashes.}
        \label{fig:populations}
    \end{figure}
    
    Fig.~\ref{fig:populations} shows the individual probabilities of finding each ground state for each problem, and Table~\ref{table1_v2} gives the total probability of finding all ground states. The blue bars in Fig.~\ref{fig:populations} demonstrate our claim that standard QAOA favors one solution more than others, while the purple bars show that G-QAOA does not. In the G-QAOA result for the edge cover problem on the triangle graph at $p=1$, the experiment closely approximates the simulation. Furthermore, we see a small improvement in the total ground state probabilities at $p=2$ compared with $p=1$ for all three graphs despite the deeper circuits at higher $p$. 
    
    Fairness describes how well the experimentally sampled distribution represents the ideal distribution. The fair sampling of ground states with equal probabilities provides a convenient method to enumerate all ground states of a problem. One measure of the efficiency of the enumeration is the average number of experimental shots required to observe each ground state at least once. To estimate this, we sample the states from the experimentally measured distributions on a classical computer. For each experimental distribution, we record the number of random draws required to observe any $N_{g}$ different ground states at least once, varying $N_{g}$ from $2$ to the total number of ground states. For each $N_{g}$ we repeat the procedure for 100,000 times to determine the average number of draws and the uncertainty. 
    
    Table~\ref{counting} shows that, at $p=1$, G-QAOA and QAOA perform similarly at enumerating ground states, with QAOA being marginally more efficient in most cases. The only exception is the paw graph problem for $N_{g}=5$, where the ground state with the smallest probability found by QAOA (the fourth ground state from the left, see Fig.~\ref{fig:populations}) has considerably lower probability in both simulation and experiment than the ground state with the smallest probability found by G-QAOA. Since the efficiency of enumerating by sampling is limited by the ground states with the lowest probability, G-QAOA shows an advantage in this case. QAOA at $p=2$ suppresses some ground states more strongly than at $p=1$, and therefore has the worst results. G-QAOA at $p=2$ shows results similar to G-QAOA at $p=1$, with slight improvement seen in a few cases. This differs from the ideal prediction that the number of draws required should decrease from $p=1$ to $p=2$ for G-QAOA since the total probability of ground states grows. However, in experiment the total probabilities do not increase significantly due to decoherence (see Table.~\ref{table1_v2}). For the same reason, we see that most often QAOA at $p=1$ requires the least number of draws to find $N_g$ ground states, while theory predicts that G-QAOA at $p=2$ should have an advantage in most cases, especially when $N_g$ is close to the total number of ground states. Nevertheless, G-QAOA at $p=1$ and $p=2$ still require fewer samples than random guessing to obtain all ground states. 

    \begin{table}[t]
        \subfloat[][Experimental data]{
                \begin{tabular}{ p{2.2cm} p{1.9cm} p{1.5cm} p{1.5cm} }
                \hline
                & &$p=1$&$p=2$\\
                \hline
                \multirow{3}{*}{unweighted}&\multicolumn{1}{l}{triangle}&5754(32)&406(8)\\ 
                &\multicolumn{1}{l}{paw}&328(4)&208(4)\\
                &\multicolumn{1}{l}{square}&602(13)&192(4)\\
                \hline
                \multirow{2}{*}{weighted}&\multicolumn{1}{l}{paw}&38(1)&90(1)\\ 
                &\multicolumn{1}{l}{square}&35(1)&45(1)\\
                \hline
                \hline
             \end{tabular}
             }
        \vspace*{5mm}
        \subfloat[][Synthetic data]{
            \begin{tabular}{ p{2.2cm} p{1.9cm} p{1.5cm} p{1.5cm} }
                \hline
                & &$p=1$&$p=2$\\
                \hline
                \multirow{3}{*}{unweighted}&\multicolumn{1}{l}{triangle}&6526(3259)&4337(943)\\ % \cline{2-4}
                &\multicolumn{1}{l}{paw}&1882(217)&4044(702)\\
                &\multicolumn{1}{l}{square}&2544(414)&2318(247)\\
                \hline
                \multirow{2}{*}{weighted}&\multicolumn{1}{l}{paw}&297(74)&1958(675)\\ 
                &\multicolumn{1}{l}{square}&674(82)&1370(270)\\
                \hline
                \hline
             \end{tabular}
             }
        \caption{Number of shots to reject the fair sampling hypothesis by re-sampling from the experimental data and the synthetic data. The synthetic data is randomly drawn from the ideal distributions. Averages and error bars are calculated by repeating the ``shots to reject" test 10 times. }
        \label{num_of_shots}
    \end{table}
    
    \begin{table}[ht!] 
    \centering
        \subfloat[][Experimental data]{
            \begin{tabular}{ p{2.2cm} p{1.9cm} p{1.5cm} p{1.5cm}}
            \hline
            &&$p=1$&$p=2$\\
            \hline
            \multirow{3}{*}{unweighted}&\multicolumn{1}{l}{triangle}&0.0007(4)&0.007(2)\\ 
            &\multicolumn{1}{l}{paw}&0.015(2)&0.016(2)\\
            &\multicolumn{1}{l}{square}&0.007(1)&0.020(2)\\
            \hline
            \multirow{2}{*}{weighted}&\multicolumn{1}{l}{paw}&0.079(8)&0.037(4)\\ 
            &\multicolumn{1}{l}{square}&0.089(9)&0.074(6)\\
            \hline
            \hline
        \end{tabular}}
    \vspace{5mm}
    \subfloat[][Synthetic data]{
    \begin{tabular}{ p{2.2cm} p{1.9cm} p{1.5cm} p{1.5cm}}
            \hline
            &&$p=1$&$p=2$\\
            \hline
            \multirow{3}{*}{unweighted}&\multicolumn{1}{l}{triangle}&0.0004(2)&0.0003(1)\\ 
            &\multicolumn{1}{l}{paw}&0.0006(3)&0.0005(2)\\
            &\multicolumn{1}{l}{square}&0.0008(3)&0.0007(3)\\
            \hline
            \multirow{2}{*}{weighted}&\multicolumn{1}{l}{paw}&0.004(2)&0.0005(2)\\ 
            &\multicolumn{1}{l}{square}&0.0021(8)&0.0009(3)\\
            \hline
            \hline
            \end{tabular}
            }
        \caption{\label{table4_KL} KL divergence between the simulation and experimental data, and synthetic data. The error bar on each experimental result is calculated by resampling from the experimental distribution 300 times, while for each synthetic result it comes from 300 sets of synthetic data generated for each problem.}
    \end{table}
    
\section{\label{sec:4}fair-sampling with G-QAOA on weighted graphs}
    In this section, we solve the edge cover problem when a numerical weight $(1-q)>0$ with $q\neq 0.5$ is assigned to each edge on the paw and square graph. The weight of each subgraph $G'$ is defined as  $P_{G'}=(1-q)^{n'}q^{n-n'}$, where $n'$ is the number of edges included in $G'$. G-QAOA samples all ground states, with squared amplitudes proportional to these weights. 
    
\begin{figure}
        \centering
        \includegraphics[width=\linewidth]{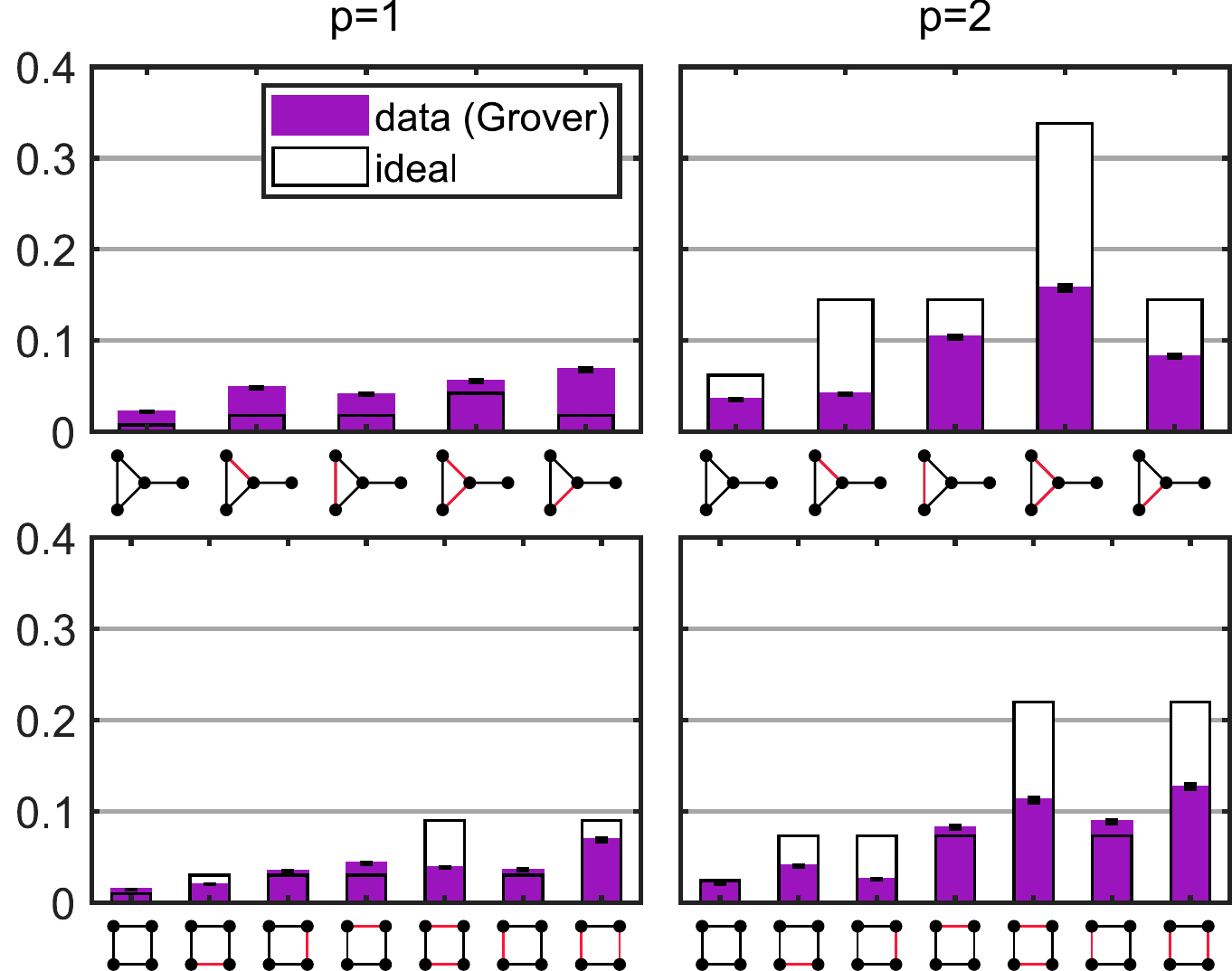}
        \caption{Simulated and experimental results from G-QAOA for the edge cover problem on the weighted paw (top) and square (bottom) graph.}
        \label{fig:weighted_populations}
    \end{figure}

    $H_{A}^{\text{ec}}$ is given in Eq.~\eqref{equ2} as in the previous two Sections. $H_{B}$ is the Grover mixer given in Eq.~\eqref{Grover}. The initial state is prepared by applying 
    \begin{equation}
        U =e^{-i\sigma^{y}\sin^{-1}{\sqrt{q}}}
        \label{U}
    \end{equation}
    to each qubit in $\ket{0}$.
    Fig.~\ref{fig:weighted_populations} shows the G-QAOA results on a paw graph with $q=0.7$ and a square graph with $q=0.75$, and compares them to the ideal population distributions. 
    
    To allow direct fairness comparisons between results for problems with different numbers of ground states and different sample sizes $N$ on weighted graphs, we need a metric that is insensitive to these problem specifics. Fairness of the sampling result can be quantified as the discrepancy between the ideal ground state distribution $Q_{2}$ and the experimentally measured ground state distribution $Q_{1}$, which is obtained by post-selecting out all ground states from 4000 experimental shots. The first method we adopt is the ``shots to reject" method, proposed in Ref.~\cite{golden2021qaoabased}, which is constructed based on the chi-squared ($\chi^{2}$) test. By re-sampling from $Q_{1}$, the goal is to compute the number of samples, $N^{*}$, needed to reject the null hypothesis $H_{0}$ at a selected significance level. Here, $H_{0}$ is that $Q_{1}$ is sampled from $Q_{2}$. The more the experimental data $Q_{1}$ deviates from the ideal distribution $Q_{2}$, the smaller $N^{*}$ will be. To find $N^{*}$, we follow the described protocol: 1) randomly draw 1000 sets of samples of size $M$ (starting from $M=2$) from $Q_{1}$, perform $\chi^{2}$ test between each sampled distribution and $Q_{2}$, and record the $1000$ p-values from the tests 2) compare the median of the p-values with the preset significance level, which is chosen to be 0.05 here, 3) if the median of the p-values exceeds the threshold, set $M:=2M$ and repeat steps 1 to 2; otherwise, if the median p-value is smaller than threshold, $H_{0}$ is rejected, then a bisection method is used to located the exact $N^{*}$ between $M$ and $M/2$. 
   
   Table~\ref{num_of_shots} shows the results from experimental data with synthetic data for comparison, where synthetic data are random samples drawn from $Q_{2}$ on a classical computer. Some ground states on the weighted graphs have very small expected populations, causing $H_{0}$ to be more easily rejected in the $\chi^{2}$ test, which contributes to the sampling results being overall less fair on the weighted than on the unweighted graphs for both experimental and synthetic data. We see a clear improvement in the fairness going from p=1 to p=2 in both weighted problems. This is due to G-QAOA boosting the probabilities of most of the ground states, including the least likely one, in each problem (see Fig.~\ref{fig:weighted_populations} and Table~\ref{table1_v2}). When generating the synthetic data with a fixed number of draws from the entire population, the ground state counts are larger for $p=2$ than $p=1$, leading to different ``shots to reject" results.

    An alternative way to characterize the differences between probability distributions is the Kullback–Leibler (KL) divergence, which is defined as
    \begin{equation}
        D_{KL}(Q_{1}||Q_{2})=\sum_{x \in X}Q_{1}(x)\frac{Q_{1}(x)}{Q_{2}(x)},
    \end{equation}
    where $X$ is the sample space. It can be intuitively understood as the information loss when we model $Q_{2}$ by $Q_{1}$, providing the distance between the two distributions.
    
    The results for the KL divergence analysis are presented in Table~\ref{table4_KL}. All trends are consistent with that seen in the ``shots to reject" analysis.

\section{\label{sec:5}Outlook}  
    In this work, we experimentally demonstrated that the standard QAOA results improve with increasing $p$ up to $p=3$ in optimization problems on arbitrary graphs, and show fair sampling results of G-QAOA up to $p=2$ on both unweighted and weighted graphs on a trapped-ion quantum computer. To push beyond these small demonstration problems, advances in fidelity and system size of the quantum hardware are crucial. Additionally, future studies on large-scale arbitrary graphs will challenge the connectivity of all hardware platforms, which highlights the importance of efficiently matching graphs and architectures. 
    
    Although G-QAOA does not show an advantage at enumerating the ground states over standard QAOA or when going to higher $p$ on unweighted graphs due to experimental noise, we do observe that the total probabilities of ground states grow with increasing $p$ in all cases. We also observe the fairness of sampling improves with increasing $p$ in the two weighted problems.  Future experiments could look at the relative advantage provided by more advanced mixers, such as the QED-inspired mixer designed for constrained flow problems~\cite{zhang2021qed}, follow general guidelines to engineer mixers that ensure solutions satisfy desired constraints~\cite{hen2016driver} or symmetries~\cite{selvarajan2021variational}, and study circuits which preserve specific physical symmetries in optimization problems in fermionic systems~\cite{gard2020efficient}.

\acknowledgements
KRAH and NML acknowledge funding by the Office of Naval Research (N00014-20-1-2695). This research was supported in part by the NSF (PHY-1430094, PHY-1848304, CMMI-2037545) and the Robert A. Welch Foundation (C-1872). AMG is supported by a Joint Quantum Institute Postdoctoral Fellowship. NML acknowledges support by the Maryland-Army-Research-Lab Quantum Partnership (W911NF1920181).

\appendix

\section{\label{A}Experimental Setup}
    The experiment is implemented on a programmable universal trapped-ion quantum computer with up to nine qubits and individual ion addressability. The native gate set includes single-qubit rotations around an arbitrary axis $\vec{n}$ in the $x$-$y$ plane of the Bloch sphere by angle $\theta$, $R({\hat n}, \theta)=e^{-i\vec{\sigma}\cdot \vec{n}\:\theta/2}$, rotations around the $z$ axis $R_z(\theta)=e^{-i\sigma_{z}\theta/2}$, and the two-qubit interaction $XX=e^{i\theta \sigma^{i}_{x}\sigma^{j}_{x}}$ between any pair of qubits for arbitrary $\theta$. The $R(\hat n,\theta)$ are realized via resonant Raman transitions with duration proportional to $\theta$. The $R_z(\theta)=e^{-i\sigma_{z}\theta/2}$ gates are classical phase advances in the laser beam controllers. The two-qubit entangling gates are implemented using the M\o lmer-S\o rensen scheme~\cite{PhysRevLett.82.1971,PhysRevA.59.R2539}, where the qubit spin states and the collective motional modes of the ion chain are coupled and decoupled via amplitude-modulated laser pulses~\cite{choi2014}. More details about the experimental setup are described in Ref.~\cite{Debnath2016}.

\section{\label{B}Circuits and parameters}
    The circuit in Fig.~\ref{fig:circuit3qGrover} shows the 3-qubit Grover mixer for unweighted problems used in the edge cover problem on the unweighted triangle graph. Fig.~\ref{fig:circuit4qGrover} shows the 4-qubit Grover mixer. Variational parameters $\alpha$ and $\beta$ for each problem are listed in Table~\ref{parameters}. 
    
    \begin{center}
        \includegraphics[width=0.48\textwidth]{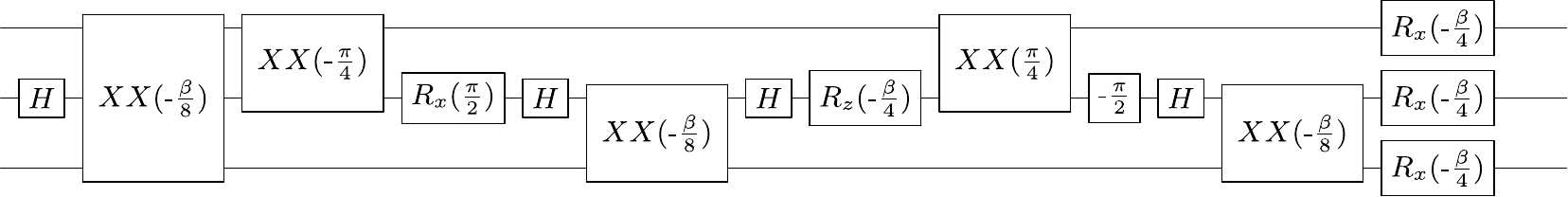}
        \captionof{figure}{\label{fig:circuit3qGrover}3-qubit Grover mixer for unweighted problems in Eq.~\eqref{Grover} with $n=3$ and $q=0.5$.}
    \end{center}
    
    \begin{center}
        \includegraphics[width=0.48\textwidth]{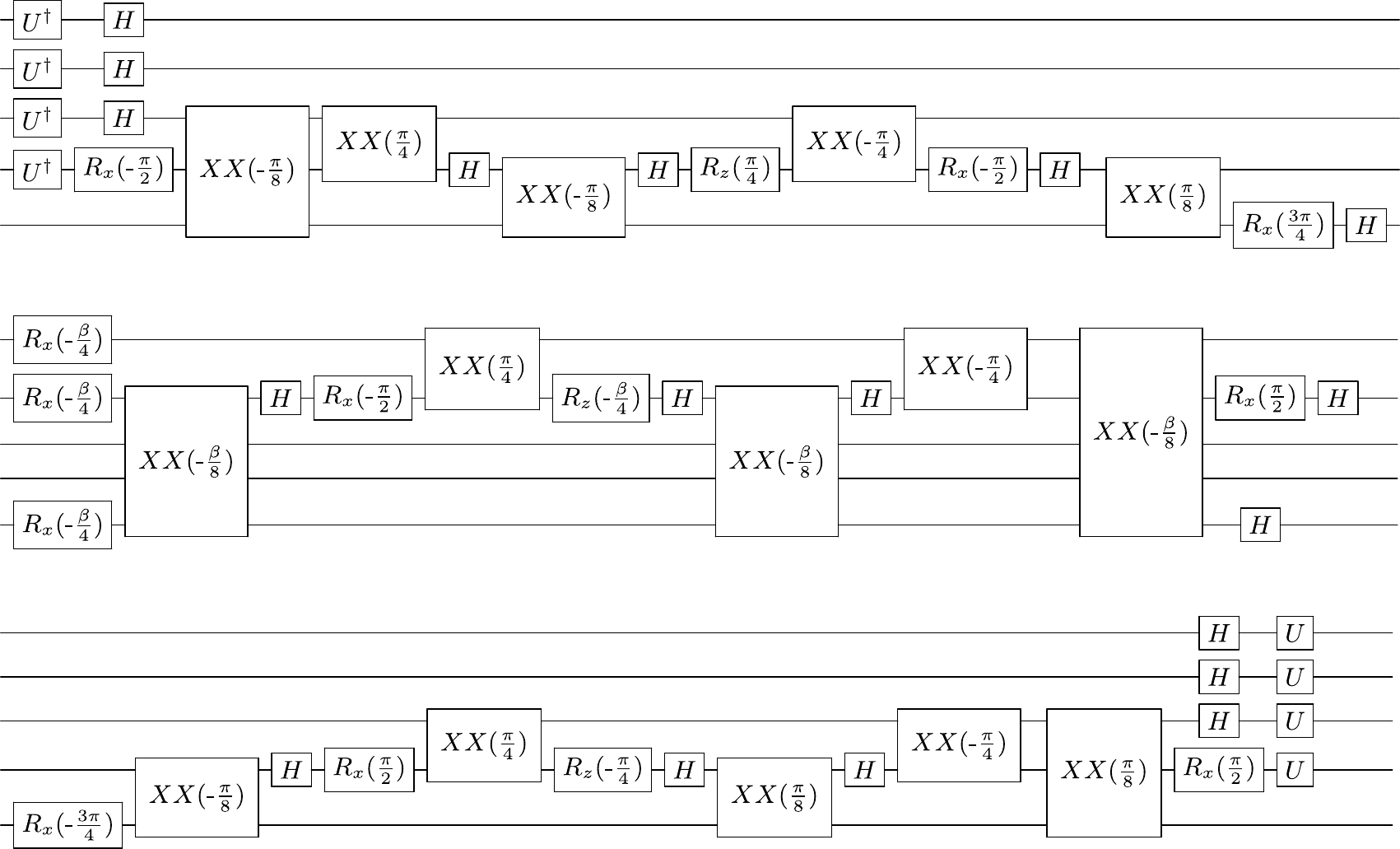}
        \captionof{figure}{\label{fig:circuit4qGrover}4-qubit Grover mixer in Eq.~\eqref{Grover} with $n=4$ and arbitrary $q$. The last qubit acts as an ancilla qubit.}
    \end{center}
    
    \begin{table*}[!htbp]
        \centering
                \subfloat[][Standard QAOA parameters corresponding to results in Fig.~\ref{fig:standardmixer} and Table~\ref{table1_v2} (left)]{
            \begin{tabular}{p{2cm} p{4cm} p{4cm} p{4cm}}
            \hline
            &$p=1$&$p=2$&$p=3$\\
            \hline
            \multirow{3}{*}{triangle}&$\alpha=-0.95$, $\beta=\:\:\:1.00$&$\alpha_1=-0.83$, $\beta_1=\:\:\:3.14$&\\
            &&$\alpha_2=\:\:\:0.85$, $\beta_2=-1.66$&\\
            % &&&\alpha_3$=$0, \beta_3$=$0\\
            \hline
            \multirow{3}{*}{square}&$\alpha=-0.87$, $\beta=-2.60$&$\alpha_1=\:\:\:0.80$, $\beta_1=-1.58$&\\
            &&$\alpha_2=-0.82$, $\beta_2=-2.28$&\\
            % &&&\alpha_3$=$0, \beta_3$=$0\\
            \hline
            \multirow{3}{*}{paw}&$\alpha=\:\:\:1.04$, $\beta=-0.61$&$\alpha_1=\:\:\:0.62$, $\beta_1=\:\:\:0.75$&$\alpha_1=\:\:\:0.5$,\:\: $\beta_1=\:\:\:1.6$\\
            &&$\alpha_2=\:\:\:0.88$, $\beta_2=-1.04$&$\alpha_2=-0.5$,\:\: $\beta_2=\:\:\:0.5$\\
            &&&$\alpha_3=-0.5$,\:\: $\beta_3=\:\:\:0.3$\\
            \hline
            \multirow{3}{*}{bridge}&$\alpha=\:\:\:0.29$, $\beta=\:\:\:0.31$&$\alpha_1=-0.55$, $\beta_1=\:\:\:0.42$&$\alpha_1=-2.09$, $\beta_1=\:\:\:0.43$\\
            &&$\alpha_2=-2.95$, $\beta_2=\:\:\:0.87$&$\alpha_2=-2.17$, $\beta_2=\:\:\:1.28$\\
            &&&$\alpha_3=-1.02$, $\beta_3=\:\:\:2.3$\\
            \hline\\
            \end{tabular}
        }
         \vspace{0.5cm}
        \subfloat[][G-QAOA parameters for unweighted problems corresponding to results in Fig.~\ref{fig:populations} and Table~\ref{table1_v2} (right)]{
            \begin{tabular}{p{1.2cm} p{3.2cm} p{3.2cm}}
            \hline
            &$p=1$&$p=2$\\
            \hline
            \multirow{2}{*}{triangle}&$\alpha=2.48$, $\beta=1.37$&$\alpha_1=0.69$, $\beta_1=1.32$\\
            &&$\alpha_2=1.22$, $\beta_2=0.92$\\
            \hline
            \multirow{2}{*}{square}&$\alpha=0.65$, $\beta=1.46$&$\alpha_1=0.48$, $\beta_1=1.52$\\
            &&$\alpha_2=0.91$, $\beta_2=0.92$\\
            \hline
            \multirow{2}{*}{paw}&$\alpha=0.79$, $\beta=1.60$&$\alpha_1=0.56$, $\beta_1=1.47$\\
            &&$\alpha_2=0.98$, $\beta_2=1.17$\\
            \hline
            \end{tabular}
        }
        \hspace{1mm}
        \subfloat[][G-QAOA parameters for weighted problems corresponding to results in Fig.~\ref{fig:weighted_populations}]{
            \begin{tabular}{p{1.2cm} p{3.2cm} p{3.2cm}}
            \hline
            &$p=1$&$p=2$\\
            \hline
            \multirow{2}{*}{square}&$\alpha=\:\:\:2.67$, $\beta=-2.30$&$\alpha_1=0.68$, $\beta_1=2.20$\\
            &&$\alpha_2=1.05$, $\beta_2=1.95$\\
            \\
            \hline
            \multirow{2}{*}{paw}&$\alpha=-2.85$, $\beta=\:\:\:2.81$&$\alpha_1=2.05$, $\beta_1=2.80$\\
            &&$\alpha_2=1.98$, $\beta_2=2.98$\\
            \\
            \hline
            \end{tabular}
        }
        \caption{ \label{parameters}Standard QAOA and G-QAOA parameters.}
    \end{table*}

% \bibliography{references}

\end{document}